\documentclass[journal]{IEEEtran}

\usepackage{amsmath,amssymb}
\usepackage{graphicx}
\usepackage{booktabs}
\usepackage[colorlinks=true,linkcolor=blue,citecolor=blue,urlcolor=blue]{hyperref}
\usepackage{orcidlink}

\graphicspath{{./}{./figs/}}

\newcommand{\bff}{f}
\newcommand{\cmark}{\checkmark}
\newcommand{\xmark}{--}

\begin{document}

\title{A Loss-Robust Disturbance Certificate for
Minimal-Receiver Quantum Key Distribution}
\author{Roberto~Di~Pietro~\orcidlink{0000-0003-1909-0336},~\IEEEmembership{Fellow,~IEEE}%
\thanks{R.~Di~Pietro is with the Computer, Electrical and Mathematical
Sciences and Engineering Division (CEMSE), King Abdullah University of
Science and Technology (KAUST), Thuwal, Saudi Arabia
(e-mail: roberto.dipietro@kaust.edu.sa).}%
\thanks{This work has been submitted to the IEEE for possible publication.}}

\markboth{}{}

\maketitle

\begin{abstract}
Quantum Key Distribution (QKD) enjoys information-theoretic security,
yet the most damaging attacks against deployed systems exploit the
receiver, where the key bit is encoded in which one of a pair of
never-identical detectors clicks. The minimal receiver, one rotatable
polarizer and one threshold detector, removes that attack surface, and
single-detector BB84 demonstrations already run sampled error
estimation; the structure of its zero-probability error subensemble,
however, has remained uncharacterized. We characterize exactly that
structure, introducing a \emph{deterministic impossible-event
certificate}: a click behind a polarizer set orthogonal to the
transmitted state has probability exactly zero on an ideal channel, so
a single occurrence is a probability-one witness of disturbance; and,
since loss deletes clicks and never creates them, the certificate is
\emph{loss-robust}. We prove it sound but incomplete over three
polarization states, and show that the four BB84 states close the gap:
a fixed-basis intercept-resend attack yields an ideal trip probability
of $1/4$ per orthogonal round ($\eta/4$ observed at detection
efficiency $\eta$), independent of the interception angle. An
illustrative finite-size budget yields $256$ retained bits from
$\approx 62{,}000$ transmitted rounds at $\eta = 0.1$; under realistic
detector noise ($q_0 = 10^{-6}$ per opened gate), each trip retains
$\approx 12$ bits of evidence at a sub-percent honest false-abort
probability per session. The core ideal trip-probability predictions
are numerically verified on the Qiskit circuit simulator, via a
released, seed-fixed implementation. 

Overall, by endowing the minimal-detector receiver of polarization QKD with a conclusive, loss-robust disturbance alarm, our solution lowers the hardware entry cost of security-monitored QKD, hence fostering its adoption at the cost-sensitive network edge.
\end{abstract}


\section{Introduction}
\label{sec:intro}

\IEEEPARstart{Q}{uantum} networks will not be built from flagship
point-to-point links but from many endpoints: access nodes, metropolitan
mesh terminals, and integrated-photonic transceivers, where per-node
receiver cost and calibration burden, rather than raw key rate, set what
is deployable at scale. Each single-photon detector carries its own
cooling, gating, and characterization overhead, so the number of
detectors per receiver is a first-order driver of network economics. This
makes the \emph{minimal} receiver, i.e., one detector, strategically
interesting for the edge tiers of a quantum network, provided it can be
made to detect eavesdropping at least as well as its multi-detector
counterparts. The obstacle has never been the hardware, which is long
established, nor error monitoring as such, which single-detector BB84
demonstrations already run by sampled estimation
\cite{martelli,singledet}; what has been missing is a detection theory
for the receiver's \emph{conclusive}, zero-probability events. In this
article we supply such a theory and experimentally show its viability.

The starting point is a well-known vulnerability. The most damaging
practical attacks on deployed quantum
key distribution do not break the protocol; they exploit the receiver.
The time-shift attack \cite{timeshift}, faked-state and blinding
attacks on the detection unit \cite{blinding}, and the broader
detector-efficiency-mismatch class \cite{effmismatch} all rest on one
architectural fact: in a standard
BB84 \cite{bb84} receiver the key bit is encoded in \emph{which} of a
pair of imperfect, never-identical detectors clicks, and every residual
difference between the paired devices is a parameter the eavesdropper
can steer.

\begin{figure*}[!t]
\centering
\includegraphics[width=\textwidth]{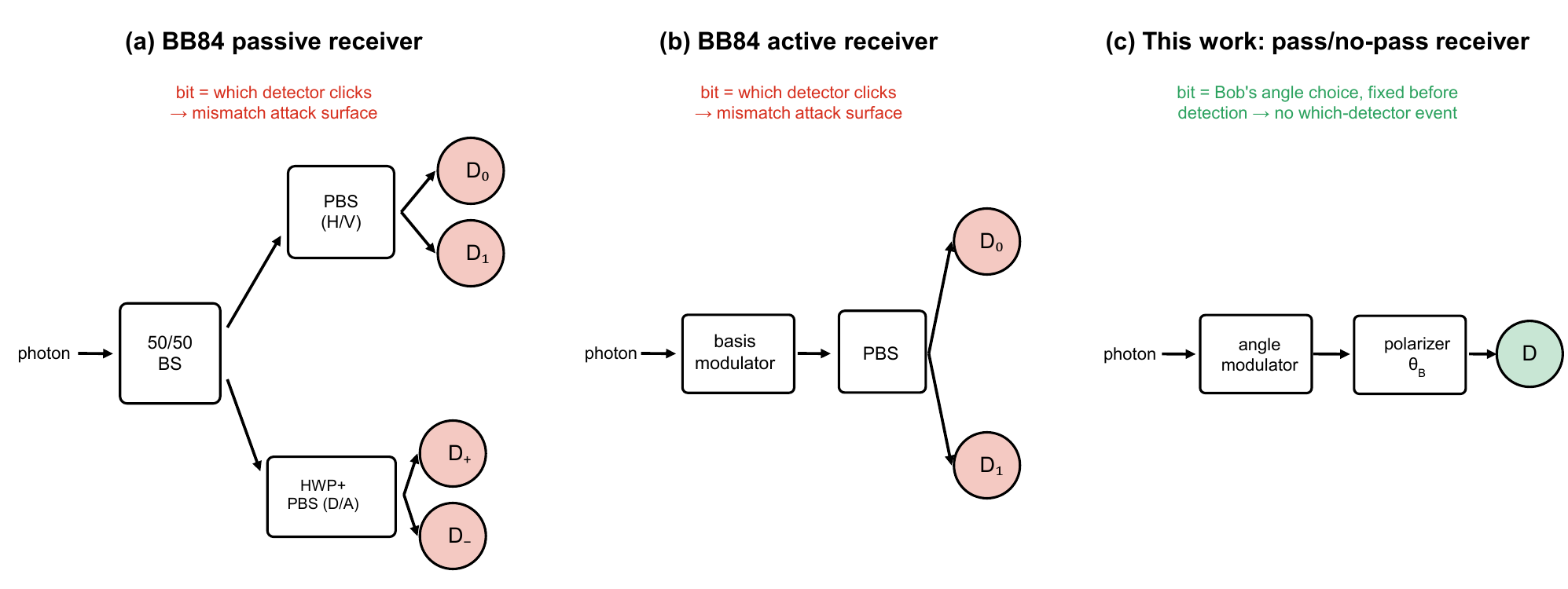}
\caption{Three receiver architectures for polarization QKD. (a) Passive
BB84: four detectors; the bit is read from which detector clicks.
(b) Active BB84: two detectors; the which-detector encoding remains,
and with it the efficiency-mismatch, time-shift, and faked-state attack
surface. (c) The single-detector pass/no-pass receiver: the bit
is Bob's angle choice, fixed before detection. Detectors are
labelled by the outcome they register: in (a) the incoming photon is split between a rectilinear (H/V) and a diagonal (D/A) analyser, whose
outputs are $D_0,D_1$ and $D_+,D_-$ respectively; in (b) the basis
modulator selects the basis before a single polarizing beam splitter,
whose two outputs are $D_0,D_1$. In both cases the sifted bit \emph{is}
the detector index, precisely the quantity an eavesdropper steers. In
(c) the single detector $D$ carries no index, because no bit is encoded
in which device fires. 
}
\label{fig:receivers}
\end{figure*}

The receiver that removes this surface is well known to be the minimal
one: a single rotatable polarizer followed by a single threshold
detector, reporting \emph{pass} or \emph{no-pass} per gated round
(Fig.~\ref{fig:receivers}c). Single-detector reception has a genuine
history. It is the optical head of practical B92 \cite{b92}
implementations \cite{b92impl}; four-setting single-SPAD BB84 itself
has been demonstrated with a variable Faraday rotator \cite{martelli}
and, recently, in decoy-state operation with a single laser and a
single detector \cite{singledet}; the time-shift literature analyzes
(and attacks) receivers that time-multiplex both bit values onto one
detector \cite{timeshift}; and,
the recently demonstrated ``four-state Bob'' countermeasure
\cite{fourstatebob}. This latter one establishes that symmetrizing logical detection
events over the physical detectors suppresses the entire
efficiency-mismatch class, explicitly including the limiting case of a
single detector. The side-channel case for one detector, in other words,
has been made.

What has \emph{not} been made is the case for the receiver's
conclusive events. With one detector, a missing click is
indistinguishable from channel loss, so any per-round test built on
absences fails; and since a basis-resolving pair extracts a bit, and
with it an error statistic, from every detected photon, practical QKD
favors detector pairs on throughput grounds, while time-multiplexed
workarounds merely relocate the which-detector side channel into a
which-time-slot side channel that the time-shift attack exploits in
its purest form \cite{timeshift}. Error monitoring itself was never
the obstacle: the single-SPAD BB84 demonstrations
\cite{peng,martelli,singledet} run sampled error estimation. Our
observation is that this estimation stops short of the strongest
reading of the same data: on this receiver, the sampled matched-basis
error event \emph{is} precisely an orthogonal pass, an event whose
ideal honest probability is zero. Being a click rather than an absence, it is immune to the
loss ambiguity that defeats missing-click tests. Reading it as a
zero-background decision event, and characterizing what that reading
buys and costs, is the purpose of this contribution.

The mechanism is a \emph{deterministic impossible-event certificate}. If
Alice transmits polarization $\theta_A$ and Bob's polarizer is set
orthogonal  $(| \theta_B - \theta_A| = 90^\circ)$, the honest click
probability is $\eta\cos^2(90^\circ) = 0$ for \emph{every} value of the
end-to-end detection efficiency $\eta$ (defined in Section~\ref{sec:receiver}). A click
on such an \emph{orthogonal round}, called a
{\bf \em trip}, is a probability-one witness that the arriving state is
not the transmitted one. Two structural properties distinguish trips
from all threshold tests. First, \emph{loss-robustness}: blocking,
attenuating, or selectively deleting photons removes clicks and can
never create one, so an observed trip cannot be forged by an adversary
manipulating loss (the classic failure mode of tests built on depressed
click rates), although suppressing the line does slow the certificate's
detection at a rate the parties observe. Second, \emph{certificate semantics}: the honest
probability is zero, not merely small, so the rule ``one trip
$\Rightarrow$ abort'' has no false-alarm penalty on an ideal channel and
each trip is individually attributable to disturbance.

\noindent\textbf{Contributions.}

What we contribute is: (i) the
identification of this ordinary error event 
as an ideal zero-background
decision event, formalized through a sampled certificate protocol with
loss-robust soundness; (ii) its completeness analysis as an abort
statistic: three states leave an eigenbasis escape, while the
four-setting average pins the ideal trip probability at $1/4$ for any
fixed-basis projective intercept-resend angle, BB84's familiar $25\%$
disturbance identity restricted to orthogonal rounds; and, (iii) a
finite-size cost estimate under the intercept-resend model (a
structural factor two versus BB84) with a false-trip correction
covering dark counts and orthogonal leakage, each trip retaining
$\approx 12$ bits of evidence at practical operating points and the
ideal predictions numerically verified on the Qiskit simulator.

\noindent\textbf{Roadmap.} Section~\ref{sec:prior} positions the contribution against the
relevant literature. Section~\ref{sec:receiver} introduces the receiver, defines the
certificate together with its sampled sifting protocol, and establishes
its loss-robustness and its soundness against arbitrary
quantum-channel attacks in the trusted-device model. Section~\ref{sec:states} shows the certificate is incomplete over three
polarization states and that the four BB84 states close the gap with an
angle-independent trip constant of $1/4$. Section~\ref{sec:qiskit}
numerically verifies the ideal trip-probability predictions on the
Qiskit circuit simulator,
including a deferred-measurement realization of the intercept-resend
adversary.
Section~\ref{sec:deploy} turns to deployment:
the photon budget, the degradation under real detectors (including dark count), and the
hardware economics. Section~\ref{sec:conclusion} concludes the paper.

\section{Related Work
}
\label{sec:prior}
In this section, we summarize the relevant work in the literature and
compare it against our solution (Table~\ref{tab:prior}).

\emph{Single-detector BB84 \cite{peng,martelli,singledet,timeshift}.}
The single-detector line goes back at least to Peng et al.'s 2007
decoy-state experiment, which already used only one detector
\cite{peng}. Four-setting single-SPAD BB84 is demonstrated: Martelli et
al.~\cite{martelli} realize it with a variable Faraday rotator, and
Donaldson et al.~\cite{singledet} extend it to decoy-state operation,
time-multiplexing all inputs onto one detector. Both run conventional
sampled error estimation; neither isolates the orthogonal
zero-probability subensemble as a per-event certificate, nor provides
the completeness, angle-blindness, or zero-count finite-size analysis
developed here --- exactly the delta this paper contributes on that
architecture. Separately, time-multiplexed designs map bit values to
time slots on one SPD. The time-shift attack applies to the latter in its sharpest
form: the bit is now encoded in \emph{when} the detector may click,
and an Eve who shifts arrival times can in principle read the full key
without errors \cite{timeshift}. This is the cautionary lesson our
receiver must answer: eliminating the detector pair is worthless if the
bit is re-encoded in another detector-observable degree of freedom. In
the pass/no-pass receiver the bit is Bob's \emph{angle choice}, fixed
before detection and never encoded in which detector, which slot, or
whether a click occurs at all, though the angle modulator inherits the
same emission-security requirements as active-BB84 basis selection, and
angle-dependent detection efficiency \cite{grasselli} must be bounded in
calibration.

\emph{Four-state Bob \cite{fourstatebob,effmismatch}.} The
bit-assignment-randomization countermeasure, recently demonstrated on a
GHz-clocked system \cite{fourstatebob}, symmetrizes logical detection
events across physical detectors and provably suppresses the
efficiency-mismatch class, including, as its authors note, in the
limiting case of a single detector. 
What \cite{fourstatebob} leaves untouched is
detection: their receiver still runs standard QBER monitoring on a
basis-resolving setup. The gap is the detection theory for the receiver
that has \emph{no} basis-resolving optics; this paper addresses that
gap.

\emph{MDI-QKD \cite{mdi}.} Measurement-device independence removes
detector trust entirely, at the cost of an untrusted Bell-state-measuring
relay: strictly more infrastructure. The single-detector receiver with
the certificate is the opposite end of the trade space: strictly less
hardware than BB84, immunity limited to the inter-detector class, and
(unlike MDI) no protection against attacks on the detector itself, such
as blinding \cite{blinding}. Both points on the spectrum are legitimate;
they serve different tiers of a quantum network, whose trust
architecture quantum resources are themselves reshaping
\cite{quantumblockchain}.

\begin{table*}[!t]
\caption{Properties by Approach: Who Provides What}
\label{tab:prior}
\centering
\setlength{\tabcolsep}{4.5pt}
\begin{tabular}{lccccccc}
\toprule
Property & BB84 & B92 & Time-mux & 1-SPAD BB84 & Four-state Bob & MDI-QKD & \textbf{This work} \\
 & \cite{bb84} & \cite{b92,b92impl} & \cite{timeshift} & \cite{peng,martelli,singledet} & \cite{fourstatebob} & \cite{mdi} & \\
\midrule
Detectors at receiver & 2--4 & 1--2 & 1 & 1 & 2 (or 1) & 2+ (relay) & \textbf{1} \\
Bit not in detector/temporal output label & \xmark & \cmark$^{e}$ & \xmark$^{a}$ & impl.-dep.$^{f}$ & \cmark & \cmark$^{b}$ & \cmark \\
Det.-eff.-mismatch class not security-critical \cite{timeshift,effmismatch} & \xmark & impl.-dep. & \xmark$^{a}$ & impl.-dep.$^{f}$ & \cmark & \cmark & \cmark$^{c}$ \\
Immune to detector-blinding \cite{blinding} & \xmark & \xmark & \xmark & \xmark & \xmark & \cmark & \xmark \\
No basis-resolving optics at receiver & \xmark & \cmark & impl.-dep. & impl.-dep.$^{f}$ & \xmark & \cmark$^{b}$ & \cmark \\
Orthogonal-pass event as abort certificate & \xmark$^{d}$ & \xmark$^{e}$ & \xmark & \xmark$^{f}$ & \xmark & \xmark & \cmark \\
Completeness analysis (3 vs 4 states) & --- & --- & --- & --- & --- & --- & \cmark \\
Angle-blind detection constant ($1/4$) & --- & --- & --- & --- & --- & --- & \cmark \\
Dark-count-corrected per-event evidence bound & --- & --- & --- & --- & --- & --- & \cmark \\
Sifted fraction (ideal, pre-sacrifice) & $\eta/2$ & $\eta/4$ & $\eta/2$ & impl.-dep.$^{f}$ & $\eta/2$ & varies & $\eta/4$ \\
Extra infrastructure beyond BB84 & --- & less & less & less & RNG bit & BSM relay & less \\
\bottomrule
\multicolumn{8}{p{0.94\textwidth}}{\footnotesize
$^{a}$Bit re-encoded in the time slot; the time-shift attack applies in
its sharpest form \cite{timeshift}.
$^{b}$Detectors sit at the untrusted relay; the users' devices need no
basis-resolving detection.
$^{c}$For the key layer: the bit is the receiver's angle choice, fixed
before detection; angle-dependent efficiency must be bounded in
calibration \cite{grasselli}.
$^{d}$BB84 with a basis-resolving receiver contains an analogous
impossible event (a matched-basis error), but its threshold test is
statistical because real channels are noisy; the event is not used as a
loss-robust certificate.
$^{e}$B92 consumes its conclusive click to generate key; in the
single-detector realization Bob's secret setting is his bit and only
pass positions are announced.
$^{f}$Runs conventional sampled QBER estimation without isolating the
zero-probability subensemble; properties marked impl.-dep.\ differ
between the four-setting \cite{martelli} and time-multiplexed
\cite{singledet} realizations.
}
\end{tabular}
\end{table*}

\emph{B92 \cite{b92}.} Single-detector B92 is the closest ancestor of
the present construction. The certificate's physics, i.e., a click that
has probability zero unless the state differs, is its operational heart
\cite{b92,b92impl}, and its announcement structure, in which Bob
reveals pass positions while his secret setting, which is his bit,
stays hidden, prefigures the sifting of Section~\ref{sec:receiver}. The decisive
difference is what the event is \emph{used for}. B92 consumes it to
\emph{generate key}; we use the matched-orthogonal round, where a click
is impossible in the honest protocol, as an \emph{abort trigger}, and
develop the properties that role requires: loss-robust soundness, the
completeness analysis (which fails for three states and is rescued by
four), the finite-size bound on the attacked fraction, and the
false-trip degradation law. 

\section{The Receiver and the Certificate}
\label{sec:receiver}

Alice prepares single photons linearly polarized at $\theta_A$, drawn
uniformly from a small public set; Bob independently draws his polarizer
angle $\theta_B$ from the same set and records pass or no-pass; the
honest pass probability is $\eta\cos^2(\theta_A-\theta_B)$, where $\eta$
denotes the link's end-to-end detection efficiency, the product of
channel transmittance and detector efficiency, i.e.\ the probability that
a transmitted photon yields a click when Bob's polarizer is aligned with
Alice's state.

A round with $|\theta_A-\theta_B| = 90^\circ$ is \emph{orthogonal}; a
pass on it is a \emph{trip}, and the honest trip probability is exactly
zero at every $\eta$. Loss-robustness follows immediately: any
adversarial or environmental strategy that only blocks, attenuates, or
deletes photons leaves the trip probability at zero --- it can suppress
evidence, never fabricate it. Statistical tests built on
\emph{depressed} click rates enjoy no such protection: they concede the
adversary slack proportional to the uncertainty in $\eta$ and her
ability to substitute disturbance for natural loss. Note that 
what is loss-\emph{independent} is the soundness of
an observed trip, which no amount of loss can forge; the
\emph{detection power} of the certificate does scale with $\eta$, and
since channel transmittance is under Eve's control, an adversary who
suppresses the line can slow detection, at the price of a
proportionally lower detection rate that Alice and Bob observe
directly. Wherever $\eta$ appears in the bounds below, it is the
detection rate estimated from the committed record, not a trusted
constant. In short: loss-independent soundness, loss-dependent
completeness. With ideal polarization optics, the leading
detector-level mechanism that can produce an honest trip is a dark
count, i.e., a click with no photon, typically caused by thermal or
electronic noise; real optics additionally introduce a finite
orthogonal leakage (polarizer extinction, residual misalignment,
drift). Both are quantified in Section~\ref{sec:deploy}.

The classical phase must be specified with care, because on a
same-angle round the angle \emph{is} the raw bit, so angles cannot
simply be broadcast. The announcements proceed as follows: (i) Bob
publicly commits to his full detection record; (ii) Alice and Bob
announce, for every round, the \emph{basis} only (rectilinear or
diagonal), while cross-basis rounds are discarded; (iii) From the same-basis
rounds, a random \emph{test subset} is drawn using public randomness
fixed only after Bob's commitment; (iv) On test rounds only, both
parties disclose their exact angles, sacrificing those bits: a
disclosed pair with $|\theta_A-\theta_B| = 90^\circ$ is a
\emph{certificate round}; error estimation runs over the whole
disclosed same-basis detected sample, aligned clicks counting as
correct outcomes and orthogonal clicks as errors; and, (v) Undisclosed
same-basis \emph{detected} rounds retain their bits as sifted key. Eve learns bases, the test set, and the sacrificed test
angles, but no retained bit. The certificate is therefore a
\emph{sampled} test: its guarantees below hold on the disclosed
subset. An orthogonal click outside the disclosed test subset
manifests as a raw-key error (Alice and Bob then hold opposite
angles); it is never individually identified as such: its population
is statistically bounded through parameter estimation, and the
resulting discrepancies are handled by error correction.

It is worth separating soundness from completeness at the outset,
because they have very different attack-model dependence.
\emph{Soundness}, meaning that a trip implies the arriving state was disturbed,
follows purely from the honest-zero property $\eta\cos^2(90^\circ)=0$ and
therefore holds against arbitrary quantum-channel attacks in the
trusted-device model, individual, collective, or coherent (the
device-side boundary of this assumption is discussed below):
irrespective of the channel attack
strategy, an observed trip is a sound witness that the received state
differs from the expected one.
The converse quantitative relation, from trip statistics to Eve's
information, is outside the present analysis. What is specific to
the intercept-resend model in the sections that follow is
\emph{completeness} and the \emph{quantitative} trip rates, that is, how often
a given attack is forced to trip. As a \emph{disturbance witness} the certificate is thus sound against
arbitrary quantum-channel attacks in the trusted-device model; the open theory
(Section~\ref{sec:conclusion}) concerns how tightly its trip statistics
bound Eve's information, the question on which full key security rests
\cite{shorpreskill}.

One boundary deserves emphasis, since the receiver is motivated by receiver
attacks. The certificate removes the inter-detector attack surface, not the
detector itself: an adversary who blinds the detector \cite{blinding} can
induce a click on an orthogonal round and so forge a trip. Loss-robustness is
a statement about the channel, and it is the trusted-device assumption that
carries it to the device. The certificate and detector-blinding
countermeasures are therefore complementary, and a deployment needs both.

\begin{figure}[!t]
\centering
\includegraphics[width=\columnwidth]{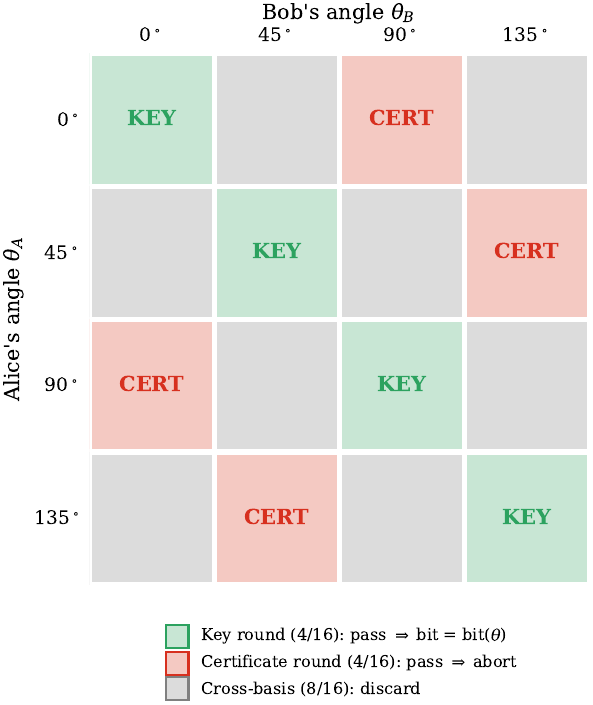}
\caption{Round classification for the four-state protocol by angle
pair $(\theta_A,\theta_B)$; exact angles are disclosed only on the
sacrificed test subset (Section~\ref{sec:receiver}). Diagonal (green):
key rounds. Orthogonal pairs (red): certificate rounds, with honest
pass probability exactly zero; one disclosed click aborts. Cross-basis
(grey): discarded. Sifted key fraction $\eta/4$ before test
sacrifice.}
\label{fig:rounds}
\end{figure}

\section{Three States Are Incomplete for the Proposed Certificate}
\label{sec:states}

\begin{figure}[!t]
\centering
\includegraphics[width=\columnwidth]{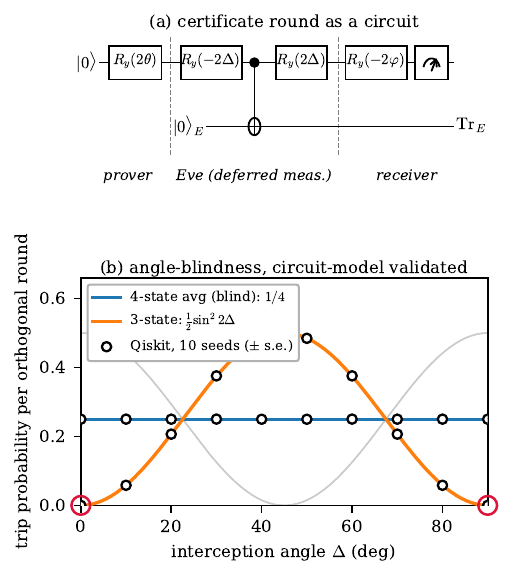}
\caption{The completeness gap, its closure, and its circuit-model
validation. (a) One certificate round as a two-qubit circuit: the prover
prepares the polarization qubit with $R_y(2\theta)$; the adversary's
intercept-resend at angle $\Delta$ is realized by deferred measurement
(an entangling gate to an ancilla in her measurement basis, discarded
via the partial trace $\mathrm{Tr}_E$), and the receiver's polarizer at $\varphi$ is a rotation followed
by one threshold measurement. (b) Trip probability versus interception
angle. Three-state scheme (orange): the trip probability
$\tfrac{1}{2}\sin^2 2\Delta$, computed over its sole orthogonal pair
$0^\circ\!\leftrightarrow\!90^\circ$, vanishes when the adversary
measures in the rectilinear eigenbasis (circled): she reads every key
bit and never trips the certificate. Four-state scheme (blue):
averaging over the four BB84 angles pins the trip probability at
exactly $1/4$ for every $\Delta$. Markers: Qiskit simulator estimates,
mean over ten seeds with standard errors smaller than the symbols.}
\label{fig:quarter}
\end{figure}

Throughout this section, trip probabilities are ideal, i.e., quoted at
$\eta = 1$; observed rates carry an additional factor $\eta$. The
result concerns the \emph{completeness of the proposed certificate},
not the security of three-state QKD in general.
With three angles $\{0^\circ,45^\circ,90^\circ\}$ (rectilinear states
carrying key, $45^\circ$ as a check), an intercept-resend adversary
measuring every photon at fixed angle $\Delta$ trips an orthogonal round
with probability $\tfrac{1}{2}\sin^2(2\Delta)$: maximal at
$\Delta=45^\circ$, but exactly \emph{zero} for
$\Delta \in \{0^\circ,90^\circ\}$
(Fig.~\ref{fig:quarter}). That eigenbasis adversary reads every
rectilinear key bit and never trips the certificate; Eve is caught
only by a statistical test on the $45^\circ$ checks, precisely the
loss-forgeable kind the certificate exists to avoid. The three-state
witness is \emph{sound} (a trip proves a disturbance) but not \emph{complete}
(a smart attack produces no trips, ever).

Keying on both BB84 bases repairs this. With the four states
$\{0^\circ,45^\circ,90^\circ,135^\circ\}$ and the same receiver
(round classes in Fig.~\ref{fig:rounds}), orthogonal rounds now exist in
\emph{both} bases, and averaging the adversary's trip probability over
Alice's four equiprobable angles makes every $\Delta$-dependent term
cancel ($\sin^2 x + \cos^2 x = 1$):

\smallskip
\noindent\fbox{\parbox{0.96\columnwidth}{\centering
\textbf{Angle-blindness.} Against a full-rate intercept-resend adversary
measuring at \emph{any} polarization angle $\Delta$, the four-state
scheme's ideal trip probability per orthogonal round is exactly $1/4$,
independent of $\Delta$; the observed probability is $\eta/4$.}}
\smallskip

No safe measurement direction remains for the adversary (Fig.~\ref{fig:quarter}); by
convexity the constant also covers adversaries who randomize their angle
round by round. Algebraically, the constant is BB84's familiar
$25\%$ intercept-resend disturbance,
$\tfrac{1}{4}\sum_\theta \tfrac{1}{2}\sin^2[2(\theta-\Delta)] =
\tfrac{1}{4}$, restricted to orthogonal rounds. The
statement assumes a state- and setting-independent detection
efficiency $\eta$ over the tested rounds; basis-dependent efficiency
must be bounded in calibration \cite{grasselli} and falls outside
it. The two mutually unbiased bases of BB84 do for the
certificate exactly what they do for the error rate, i.e., eliminate every
safe direction, but through an ideal-channel zero-false-positive event class instead
of a threshold.

Interception also leaves the standard fingerprints elsewhere: key rounds
show an error rate of $\bff/4$ for attack fraction $\bff$, identical
to BB84's intercept-resend signature, and the adversary's information
is at most $\bff/2$ bits per sifted bit. Here her optimum is an eigenbasis of
the signal states; the intermediate ``Breidbart'' direction at
$22.5^\circ$, often considered in BB84 eavesdropping analyses because
it maximizes Eve's guessing probability, gains nothing here:
angle-blindness pins her trip probability at $1/4$ whatever she
chooses.

\emph{Relation to sampled error estimation.} The certificate reads a familiar
observable in an unfamiliar regime, and the relation is worth stating exactly.
On this receiver a disclosed orthogonal click \emph{is} a matched-basis bit
error, so the certificate statistic coincides with the error count of sampled
parameter estimation on a four-setting single-detector BB84 receiver, and the
$1/4$ constant is the round-restricted form of the classical $25\%$
intercept-resend error rate. What the certificate supplies is the analysis of
that statistic's \emph{zero-count regime}, whose operational content is
threefold. First, the null is nuisance-parameter-free in the ideal model: the
honest trip probability is zero regardless of $\eta$, alignment, or channel
loss, where a QBER threshold must budget channel noise, so the false-alarm
rate is set by the device background $q_0$ rather than by a channel-noise
budget. Second, the rule is a natural \emph{sequential} alarm, aborting at the
first disclosed trip with a false-alarm budget $n_\perp q_0$ that accumulates
transparently, rather than a block-wise threshold evaluated after the fact.
Third, each disclosed click is individually attributable and carries
quantifiable evidence (Fig.~\ref{fig:bayes}), which is what makes a single
event actionable at the network layer. The two views are one statistic in two
regimes: when disclosed errors are observed, standard estimation applies and
is tighter (under intercept-resend, an observed $Q$ bounds
$\bff \approx 4Q$); when the disclosed count is zero, the zero-count bound of
Section~\ref{sec:deploy} governs. We claim the zero-count analysis, the
completeness result above, and the false-trip law below; the underlying event
class is BB84's.

\section{Numerical Verification on the Qiskit Circuit Model}
\label{sec:qiskit}

The ideal trip-probability predictions of the previous sections are
analytic, but each one also translates into an executable statement
about a small quantum circuit, so we verified them numerically on the
Qiskit simulator. A noiseless two-qubit
simulation carries no security weight of its own; its value is as an
independent, executable check that the implementation and the analysis
agree, with every number regenerable from the released script.
One certificate
round compiles to the two-qubit circuit of Fig.~\ref{fig:quarter}(a).
The prover's photon at polarization $\theta$ is the qubit
$R_y(2\theta)\,|0\rangle$; the receiver's polarizer at angle $\varphi$
is the rotation $R_y(-2\varphi)$ followed by a single computational-basis
measurement whose outcome $0$ is a click, reproducing the honest pass
probability $\eta\cos^2(\theta-\varphi)$ of Section~\ref{sec:receiver}
exactly, with the orthogonal setting landing on the zero; the
simulation runs at $\eta = 1$, where all ideal probabilities live.

The one modeling subtlety is the adversary. A literal intercept-resend
requires a mid-circuit measurement and a state re-preparation conditioned
on its result; we avoid both through deferred measurement. The adversary
rotates into her basis, entangles the signal with an ancilla through a
single controlled-NOT, and rotates back; tracing the ancilla out dephases
the signal in her measurement basis, which is, on the receiver's
marginal, statistically identical to measuring at $\Delta$ and re-emitting
the observed eigenstate. The whole attack is thus one extra qubit and
three gates, and plain shot sampling applies end to end.

The runs are seed-fixed (ten independent simulator seeds, $4\times
10^4$ shots per configuration) and reproduce every headline number.
On honest orthogonal rounds, zero trips occurred in $1.6$ million
orthogonal-round trials across all four BB84 states: the impossible
event did not happen once, at the circuit level, exactly as the
ideal-channel analysis demands. Under full-rate fixed-basis projective
intercept-resend (the model of Section~\ref{sec:states}, extended to
random basis mixtures by convexity), the four-state trip probability
came out flat at $0.2500$ across the entire
interception-angle grid, with a worst-case deviation from $1/4$ below
$3\times10^{-4}$ and ten-seed standard errors below $5\times10^{-4}$.
For the three-state set, only $0^\circ\!\leftrightarrow\!90^\circ$
forms an orthogonal pair, so test rounds exist for the rectilinear
states alone; the simulated trip probability traced the predicted
$\tfrac{1}{2}\sin^2 2\Delta$ and, at the eigenbasis
$\Delta\in\{0^\circ,90^\circ\}$, produced \emph{zero} trips: the
completeness failure of Section~\ref{sec:states}, reproduced
event-by-event, that motivates the fourth state.
Fig.~\ref{fig:quarter}(b) overlays the sampled estimates on the analytic
curves; the markers sit on the lines to within symbol size. To ensure reproducibility, the
simulation script, with all seeds declared, is released alongside the
paper.

\section{Deployment: Photons, Hardware, and Real Detectors}
\label{sec:deploy}

In this section, we assess the deployment cost of the proposed solution. In particular, we consider: (i) the photon budget; (ii) the degradation induced by real detectors; and, (iii) the hardware bill.
All three stem from a single structural fact, long known
as the reason why practical QKD adopted detector pairs: a matched-angle
\emph{no-pass} is indistinguishable from loss and must be discarded,
whereas a basis-resolving receiver extracts a bit from every detected
photon. As the following shows, detector quality
drives both the price of the receiver and the sharpness of the alarm it
hosts.

\begin{figure}[!t]
\centering
\includegraphics[width=\columnwidth]{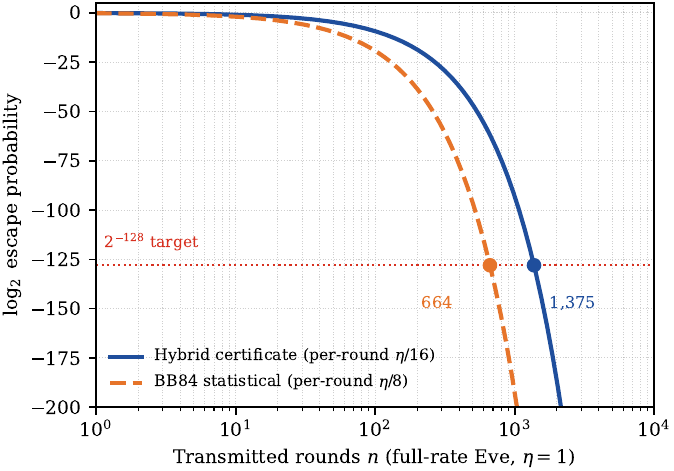}
\caption{Detection budget against a full-rate fixed-basis
intercept-resend adversary at $\eta=1$ in dedicated-test operation ($\tau=1$): trips arrive at
$\eta/16$ per transmitted round vs BB84's $\eta/8$, and $2^{-128}$
escape probability costs $1{,}375$ rounds vs BB84's $665$. At
test fraction $\tau$ the escape probability becomes
$(1-\tau\eta/16)^N$. The factor of two buys the event class:
loss-robust, zero ideal-channel false positives.}
\label{fig:escape}
\end{figure}

\emph{Photons.} The test fraction $\tau$ of Section~\ref{sec:receiver}
splits the budget: of $N$ transmitted rounds, the disclosed certificate
rounds number $n_\perp = \tau N/4$, while the retained raw key is
$n_{\rm raw} \approx (1-\tau)\eta N/4$ detections, versus BB84's
$\eta N/2$ before its own test sacrifice; a full-rate attack trips at
rate $\tau\eta/16$ per transmitted round. Against a partial-rate
adversary, zero trips over the $n_\perp$ disclosed rounds bound the
attacked fraction at $\bff \le 4\ln(1/\varepsilon)/(\eta\,n_\perp)$
with confidence $(1-\varepsilon)$, assuming attacked rounds are placed
independently of the post-commitment test selection (a deterministic
attacked fraction leads to the hypergeometric analogue with the same
leading behavior); here $\varepsilon$ is the
confidence parameter of this intercept-resend bound, not a composable
secrecy parameter, since the general-attack connection is open
(Section~\ref{sec:conclusion}). The accounting mirrors standard
finite-key practice \cite{tomamichel} while remaining a finite-size cost
estimate under the intercept-resend model rather than a composable
finite-key proof. A worked example at $\tau=1/2$,
$\eta=0.1$, $\varepsilon=2^{-64}$: $N=62{,}000$ rounds give
$n_\perp = 7{,}750$, hence $\bff \le 0.23$, and
$n_{\rm raw} \approx 775$. On this receiver the honest retained-key
error rate is itself set by the background,
$Q \approx q_0/\eta \approx 10^{-5}$; we nonetheless budget error
correction at a conservative $Q = 2\%$ allowance, an independently
budgeted quantity, not derived from the zero-trip bound (which alone
would admit an attack-induced error rate up to
$\bff/4 \approx 5.75\%$). The key length then
follows from
$\ell = n_{\rm raw}\,[1 - h(Q) - \bff/2] - \Delta_{\rm fin}$,
with $h$ the binary entropy ($h(0.02)\approx 0.14$), $\bff/2$ the
privacy-amplification rate against the certified bound (for the
fixed-basis intercept-resend family, Eve's information averaged over
the two BB84 bases is at most half a bit per attacked sifted bit,
supporting $I_E \le \bff/2$), and
$\Delta_{\rm fin}\approx 300$ bits, a conservative allowance covering
the $2\log_2(1/\varepsilon)=128$-bit privacy-amplification failure
term, error-correction verification, and smoothing corrections at
$\varepsilon=2^{-64}$ \cite{tomamichel}, yielding $\ell \approx 775\times 0.744 - 300 \approx 277
\ge 256$. The operating point is illustrative, and $n_\perp$, $n_{\rm raw}$ are
expectations; a guaranteed block additionally needs concentration on
the detection counts. The budget closes for error-correction
allowances up to $Q \approx 2.5\%$; beyond that, $N$ scales up
accordingly. The $2^{-128}$ certification of Fig.~\ref{fig:escape} is the
miss probability of a specified full-rate attack in dedicated-test
operation ($\tau=1$); at $\tau=1/2$ the same confidence costs twice
the rounds.

\begin{figure}[!t]
\centering
\includegraphics[width=\columnwidth]{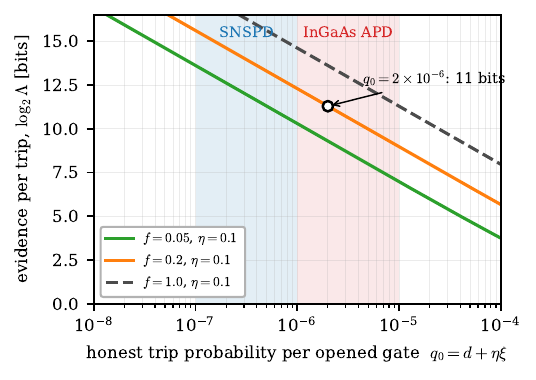}
\caption{The exact price of realism. Dark counts and orthogonal
leakage (honest trip probability $q_0 \simeq d + \eta\xi$) convert
the probability-one witness into a Bayes-factor witness of $\log_2\Lambda$ bits of evidence per trip,
$\Lambda = 1+(1-q_0)\bff\eta/4q_0$ (exact under the
independent-background model), roughly 8--14 bits over most of the
shaded detector regimes (bands mark each
detector class's dark-count range; leakage $\eta\xi$ shifts the
operating $q_0$ rightward). The honest false-abort
probability at the operating point of Section~\ref{sec:deploy}
($\tau=1/2$, $n_\perp=7{,}750$) is $\approx 0.8\%$ per session at $q_0=10^{-6}$
under the single-trip rule, and below $10^{-4}$ under a two-trip
rule.}
\label{fig:bayes}
\end{figure}

\emph{Dark counts and orthogonal leakage.} Two mechanisms can forge a
trip on a real device: a dark count inside an orthogonal-round gate,
with per-gate probability $d$ (illustrative ranges: ns-gated SNSPDs
$10^{-7}$--$10^{-6}$; InGaAs APDs $10^{-6}$--$10^{-5}$), and residual transmission through
the nominally orthogonal polarizer (finite extinction ratio,
misalignment, uncompensated drift), with effective leakage $\xi$ per
arriving photon. We define $q_0$ operationally as the calibrated
honest trip probability per opened gate,
$q_0 = P(\mathrm{trip} \mid \mathrm{honest\ device})$, with
$q_0 \simeq d + \eta\xi$ as its hardware-motivated first-order
decomposition, further honest mechanisms being absorbed into the
measured value. At illustrative operating points, rather than
representative specifications: with $\eta=0.1$ and a well-aligned
Glan-type polarizer at $\xi = 10^{-5}$, the two terms are comparable,
and at $\xi = 10^{-4}$ the leakage term dominates. On a real device the
ideal probability-one certificate is thus replaced by a statistical
witness: under the honest hypothesis a trip occurs with probability
$q_0$ per opened gate, whereas an attack adds an excess trip rate. The
false-abort probability under the single-trip rule is at most
$n_\perp q_0$, where $n_\perp$ counts \emph{opened orthogonal
gates}.
At the operating point above ($\tau = 1/2$,
$n_\perp = 7{,}750$) this is $\approx 0.8\%$ per session at
$q_0 = 10^{-6}$ and $\approx 1.5\%$ at $q_0 = 2\times10^{-6}$; where
this is unacceptable, aborting on the second trip in a session drops it
below $10^{-4}$ at $q_0=10^{-6}$ (and to $\approx 1.2\times10^{-4}$ at
$q_0=2\times10^{-6}$), at negligible cost to genuine-attack
sensitivity, since
a real interception trips each opened orthogonal gate with probability
$\eta\bff/4 \gg q_0$. Conceptually, each trip carries a
likelihood ratio of $\Lambda = 1 + (1-q_0)(\bff\eta/4)/q_0$ for the
attack hypothesis, exact under the independent-background composition
$q_1 = q_0 + (1-q_0)\bff\eta/4$ (plotted in Fig.~\ref{fig:bayes}),
reducing to
$(\bff\eta/4)/q_0$ when $\bff\eta/4 \gg q_0$, about
$2^{12}$, i.e.\ 12 bits of evidence per trip at $\bff=0.2$,
$\eta=0.1$, $q_0=10^{-6}$ (Fig.~\ref{fig:bayes}). 
The per-trip evidence scales with $\bff$, so a cautious low-rate
adversary faces the certificate from two sides at once: she trips rarely
(rate $\propto\bff$) and, when she does, each trip deposits less
evidence. This is precisely the regime in
which the attack-fraction bound $\bff \le 4\ln(1/\varepsilon)/(\eta n_\perp)$
already certifies that few key bits have leaked, so the two layers cover
complementary ranges of $\bff$ rather than leaving a gap. At very large scales ($n_\perp q_0 \gtrsim 1$) the
single-trip rule yields to a small Poisson threshold, beneath which an
adversary can hide an attack fraction comparable to the background
scale set by $q_0$ and the chosen threshold; at such scales the
certificate is a high-Bayes-factor statistical test rather than a
literal proof.

\emph{Drift and calibration.} Polarization drift in deployed fiber and
angle-dependent detection efficiency of the modulator--detector chain
\cite{grasselli} enter the same budget through $\xi$: active
compensation keeps the leakage term at the extinction-ratio floor
rather than letting drift inflate it.

\emph{Hardware.} The factor of two in photons buys the
minimal-detector receiver in polarization QKD; the economics below
assume single-photon sources, with weak-coherent decoy-state operation
left to future work. With InGaAs-APD technology each detector
channel is an independently packaged module with its own cooling and
quenching, so it dominates the per-channel bill: dropping three of
four channels (versus passive BB84) or one of two (versus active BB84)
removes the corresponding modules outright, and the saving scales with
the module count. With SNSPDs a shared cryostat dominates the receiver
bill and the marginal cost of an extra channel is small, so the cost
advantage of the minimal receiver shrinks accordingly.

Overall, the cost advantage is
strongest exactly where dark counts are worst, i.e., with APDs, and
smallest, though still present, where the certificate is cleanest,
i.e., with SNSPDs.

\section{Conclusion and Future Work}
\label{sec:conclusion}

In this paper, we have introduced a loss-robust disturbance
certificate for the single-polarizer, single-detector QKD receiver. A click behind a
polarizer set orthogonal to the transmitted state is impossible on an
ideal channel, so a single trip proves disturbance and, unlike every
test built on missing clicks, no amount of loss can forge one. We showed that three
signal states leave the witness sound but incomplete; the four BB84
states close the gap exactly, pinning the ideal trip probability at
$1/4$ per orthogonal round ($\eta/4$ observed) against an
intercept-resend adversary measuring in any polarization basis. The
guarantee costs a factor of two in photons and degrades gracefully
under real detectors into a witness still worth roughly a dozen bits of
evidence per trip. The introduced overhead is negligible: at the illustrative operating
point ($\eta = 0.1$, $\tau = 1/2$), $\approx 62{,}000$ transmitted
rounds yield a budget exceeding $256$ retained bits under the
specified intercept-resend model, i.e., some $62$~ms of integration at
a megahertz clock and $62~\mu$s at a gigahertz, while a
full-rate fixed-basis intercept-resend adversary is certified at the
$2^{-128}$ level after
$\approx 1{,}375$ rounds of dedicated-test operation at unit detection
efficiency, under two milliseconds. At $q_0 = 10^{-6}$, the honest
false-abort probability is below one percent per session under the
single-trip rule and below $10^{-4}$ if one waits for a second trip. Moreover, in terms of CapEx (a key metric for real
deployment), removing all but one detector reduces the receiver bill,
the more so where detectors are independently packaged and cooled, as
with InGaAs APD modules. 

Overall, the presented certificate endows the
minimal-detector receiver of polarization QKD with a conclusive,
loss-robust disturbance alarm at a quantified and modest cost, lowering
the hardware entry barrier of security-monitored QKD and hence
fostering its adoption at the cost-sensitive edge tiers of quantum
networks.

Three directions remain open. First, while soundness holds against arbitrary
quantum-channel attacks in the trusted-device model
(Section~\ref{sec:receiver}), the certificate layer's \emph{quantitative}
security, that is, the trip rate a given adversary is forced to incur, is
established here only for intercept-resend; no theorem yet bounds a general
collective or coherent adversary. The conjecture is structural. Writing Eve's
action as a completely positive map $\mathcal{E}$, zero orthogonal-round trips
in the rectilinear basis force $\langle\psi^\perp|\mathcal{E}(|\psi\rangle
\langle\psi|)|\psi^\perp\rangle = 0$ on the rectilinear signal states, and
simultaneously in the diagonal basis for their conjugates; a map whose output
has zero overlap with the orthogonal complement of \emph{both} mutually
unbiased signal sets should act as the identity channel on the signal-qubit
subspace, which is exactly the class that leaks no information. Making this
quantitative, by trading a small trip rate for a bound on Eve's Holevo
information through a continuity argument on that operator constraint, would
raise the witness from a detection heuristic to a security primitive, and
connects naturally to the complementarity-based security proofs
\cite{shorpreskill} in which phase-error and bit-error monitoring play dual
roles. The substantive difficulty is loss and postselection: Eve's action must
be modelled as a quantum instrument with click and loss branches, and the
theorem must show that conditioning on detection cannot leave her correlated
with the retained bits while all detectable disturbance is pushed into erasure
events. Second, single-photon sources are assumed; a weak-coherent
implementation would require a decoy-state extension, which the certificate's
per-round structure should admit but which we do not develop here. Third, an
abort signal with no ideal-channel false positives is attractive at the network
layer, where it composes with key-management orchestration without the
threshold calibration that per-link error monitoring imposes; quantifying that
composition across a multi-node topology is left to future work.

\begin{IEEEbiography}[{\includegraphics[width=1in,height=1.25in,clip,keepaspectratio]{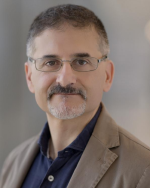}}]{Roberto Di Pietro}
(Fellow, IEEE) is currently a Full Professor of computer science with the
CEMSE, KAUST, Saudi Arabia. Previously, he was a Professor of
cybersecurity and a Founder with the CRI-Laboratory, HBKU-CSE, Doha,
Qatar. He also served as the Global Head for Cybersecurity Research for
Nokia Bell Labs. He has been working in the cybersecurity field for more
than 25 years, leading technology-oriented and research-focused teams in
the private sector, government, and academia (MoD, United Nations HQ,
EUROJUST, IAEA, and WIPO). Besides being involved in M\&A of
startups, and having founded one (exited), he is a board member of a few
research institutions. His research interests include AI-driven
cybersecurity, distributed systems security, wireless security, online
social networks security, cryptocurrencies technology, and cloud
security. He is a member of Academia Europaea and a Distinguished
Scientist of ACM. From 2011 to 2012, he was a recipient of the Chair of
Excellence from the University Carlos III, Madrid. In 2020, he was a
recipient of the Jean-Claude Laprie Award for significantly influencing
the theory and practice of Dependable Computing.
\end{IEEEbiography}


\begin{thebibliography}{15}

\bibitem{bb84}
C.~H. Bennett and G.~Brassard, ``Quantum cryptography: Public key
distribution and coin tossing,'' in \emph{Proc. IEEE Int. Conf.
Computers, Systems and Signal Processing}, Bangalore, India, 1984,
pp. 175--179.

\bibitem{b92}
C.~H. Bennett, ``Quantum cryptography using any two nonorthogonal
states,'' \emph{Phys. Rev. Lett.}, vol.~68, no.~21, pp. 3121--3124,
1992.

\bibitem{b92impl}
H.~Ko, B.-S. Choi, J.-S. Choe, and C.~J. Youn, ``Advanced unambiguous
state discrimination attack and countermeasure strategy in a practical
B92 QKD system,'' \emph{Quantum Inf. Process.}, vol.~17, no.~1,
art. 17, 2018.

\bibitem{martelli}
P.~Martelli, M.~Brunero, A.~Fasiello, F.~Rossi, and M.~Martinelli,
``Effective single-SPAD implementation of quantum key distribution,''
in \emph{Proc. Quantum Information and Measurement (QIM) V}, Optica,
2019, paper T5A.57.

\bibitem{timeshift}
B.~Qi, C.-H.~F. Fung, H.-K. Lo, and X.~Ma, ``Time-shift attack in
practical quantum cryptosystems,'' \emph{Quantum Inf. Comput.}, vol.~7,
no.~1\&2, pp. 73--82, 2007.

\bibitem{singledet}
A.~Tello~Castillo, C.~Simmons, and R.~Donaldson, ``Experimental
demonstration of polarization-based decoy-state BB84 quantum key
distribution utilizing a single laser and a single detector,''
\emph{Opt. Lett.}, vol.~50, no.~4, pp. 1184--1187, 2025.

\bibitem{peng}
C.-Z. Peng \emph{et al.}, ``Experimental long-distance decoy-state
quantum key distribution based on polarization encoding,'' \emph{Phys.
Rev. Lett.}, vol.~98, art. 010505, 2007.

\bibitem{effmismatch}
C.-H.~F. Fung, K.~Tamaki, B.~Qi, H.-K. Lo, and X.~Ma, ``Security proof
of quantum key distribution with detection efficiency mismatch,''
\emph{Quantum Inf. Comput.}, vol.~9, no.~1\&2, pp. 131--165, 2009.

\bibitem{fourstatebob}
B.~J. Taylor et al., ``Practical countermeasure against attacks
exploiting detection-efficiency mismatch in quantum key distribution,''
\emph{Phys. Rev. Applied}, 2026, doi:10.1103/7zhl-5vv1.
arXiv:2605.22580. Last Accessed: August~22, 2026.

\bibitem{grasselli}
F.~Grasselli et al., ``Quantum key distribution with basis-dependent
detection probability,'' \emph{Phys. Rev. Applied}, vol.~23,
art. 044011, 2025.

\bibitem{blinding}
L.~Lydersen, C.~Wiechers, C.~Wittmann, D.~Elser, J.~Skaar, and
V.~Makarov, ``Hacking commercial quantum cryptography systems by
tailored bright illumination,'' \emph{Nat. Photonics}, vol.~4,
pp. 686--689, 2010.

\bibitem{mdi}
H.-K. Lo, M.~Curty, and B.~Qi, ``Measurement-device-independent quantum
key distribution,'' \emph{Phys. Rev. Lett.}, vol.~108, art. 130503,
2012.

\bibitem{shorpreskill}
P.~W. Shor and J.~Preskill, ``Simple proof of security of the BB84
quantum key distribution protocol,'' \emph{Phys. Rev. Lett.}, vol.~85,
no.~2, pp. 441--444, 2000.

\bibitem{tomamichel}
M.~Tomamichel, C.~C.~W. Lim, N.~Gisin, and R.~Renner, ``Tight finite-key
analysis for quantum cryptography,'' \emph{Nat. Commun.}, vol.~3,
art. 634, 2012.

\bibitem{quantumblockchain}
Z.~Yang, T.~Salman, R.~Jain, and R.~Di~Pietro, ``Decentralization using
quantum blockchain: A theoretical analysis,'' \emph{IEEE Trans. Quantum
Eng.}, vol.~3, art. 4100716, pp. 1--16, 2022.

\end{thebibliography}
\end{document}